# Slow photon delay and the neutrino velocity

Gustavo R. González-Martín

Departamento de Física, Universidad Simón Bolívar,

Apartado 89000, Caracas 1080-A, Venezuela.

Webpage: http://prof.usb.ve/ggonzalm/

Abstract

Starting from the coordinate system used by Einstein to find the bending of light rays by gravitational fields we calculate the effect of the Earth gravitational energy along a hypothetical photon null path on the geoid non inertial system. There is an energy term, relative to an inertial system, which may be interpreted as a small time-relative "dressed" physical rest mass correction to the photon null mass. This relative gravitational potential energy determines an equivalent proper time delay proportional to the laboratory non inertial flying time interval along the trajectory with a small factor of 5.276x10-5. We also use a geometric technique based on Ehresmann connections to compare the parallel transport of the photon vector and the neutrino spinor on null trajectories. The result is that there is a theoretical neutrino time delay from its inertial path which is half the delay of the photon. Applying these delays to hypothetical particle trajectories from the CERN SPS/CNGS target to the LNGS OPERA detector we find that the difference between the experimental and theoretical results falls within the reported experimental errors. The delayed neutrino is faster than the delayed photon but its speed is smaller than the fundamental constant $c$. The results also indicate that neutrino trajectories, as photon trajectories, are affected by gravitational fields but with smaller effects.

# Introduction

In order to measure the speed of a very fast particle, for example a neutrino [1], travelling between two faraway points we would essentially compare its speed with the speed of light between these two points. This is so because time propagation and clock synchronization of a global time system is done by light-ray techniques. If the points are on the Earth geoid we must then face the problem that any reference system attached to the Earth is not an inertial frame. In such an experiment the hypothetical light speed will not be the established value $c$ which only holds for an inertial system in vacuum.

Global time systems [2, 3] rely on the "Temps Atomique International" (TAI) which is based on the measurement of the proper times of a set of clocks located on the earth geoid. The clock rates, which have the required precision, are corrected for both the relativistic kinematical effect of the Earth rotation and the relativistic dynamic effect of the Earth gravitation with the objective of providing a civil global time such as UT1 and UTC with appropriate accuracy for civil purposes. There is a weekly updated time correction DUT1 which is the reported difference between UT1 and UTC. UTC is maintained via leap seconds using DUT1.

There is another time problem in this measurement which is not associated with the TAI clock rates but directly with the trajectory. There is a photon delay on the null light trajectory relative to the total trajectory time measured by the non inertial system clocks. This long distance relativistic effect does not affect directly the TAI clocks which are the laboratory non inertial clocks but is present on the photon clock. One way to measure this uncorrected delay would be to send one light ray directly from A to B, but this is not possible by obvious reasons. A second approach would be to apply a theoretically calculated correction to the established UTC time difference between points A and B, applicable only to the speed measurement experiment. We discuss the second approach to the problem starting from the coordinate system used by Einstein to find the bending of light rays by gravitational fields [4].

## The Einstein tangent frame.

Einstein's light deflection null geodesic in a non inertial system, with the notation indicated in the appendix, is [4]

$$ds^2 = -\left(1 + \frac{\kappa}{4\pi}\int\frac{\sigma dV}{r}\right)\left(dx_1^2 + dx_2^2 + dx_3^2\right) + \left(1 - \frac{\kappa}{4\pi}\int\frac{\sigma dV}{r}\right)dl^2 = 0 \quad . \quad (1)$$

The light trajectory in the non-inertial system is

$$\left(1 + \frac{\kappa}{4\pi}\int\frac{\sigma dV}{r}\right)\left(dx_1^2 + dx_2^2 + dx_3^2\right) = \left(1 - \frac{\kappa}{4\pi}\int\frac{\sigma dV}{r}\right)dl^2 \quad .$$

We define the small energy term relative to an inertial system by

$$\varepsilon^2 = 1 - \frac{dx_1^2 + dx_2^2 + dx_3^2}{dl^2} = 1 - \frac{1 - \frac{\kappa}{4\pi}\int\frac{\sigma dV}{r}}{1 + \frac{\kappa}{4\pi}\int\frac{\sigma dV}{r}} \approx (2)\frac{\kappa}{4\pi}\int\frac{\sigma dV}{r} \quad ,$$

$$c' = \frac{dx_1^2 + dx_2^2 + dx_3^2}{dl^2} = 1 - \varepsilon^2 \quad . \quad (2)$$

The use of a modified finite non inertial time $l'$ to calculate this gravitational effect on the light trajectory is not appropriate. We should make the calculation relative to the local instantaneous tangent inertial frames or moving frames. We rearrange eq. (1) to compare with locally inertial tangent spaces. The result is

$$ds^2 = -\left(dx_1^2 + dx_2^2 + dx_3^2\right) + dl^2 - \left(\frac{\kappa\left(dx_1^2 + dx_2^2 + dx_3^2 + dl^2\right)}{4\pi}\int\frac{\sigma dV}{r}\right) = 0 \quad ,$$

$$ds^2 = -dS^2 + dl^2 - \varepsilon^2\left(\frac{dS^2 + dl^2}{2}\right) = 0 \quad (3)$$

where

$$dx_1^2 + dx_2^2 + dx_3^2 \equiv dS^2 \approx dl^2 \approx \frac{dS^2 + dl^2}{2} \equiv d\xi^2 \ ,$$

The $dl$ length-time is related to a laboratory clock flying time $dt_c \approx d\xi$ assuming light speed $c$, up to differences of order $\varepsilon^2$.

The light speed given by eq. (2) is not a scalar and it is unreliable for time calculations. Instead we may use the projection $l^0$ of the light vector on a static standard clock vector with respect to an observer, which corresponds to a direct time measurement. As indicated in the appendix, the projections $l^\alpha = \theta^\alpha_\mu l^\mu$ onto an Einstein orthogonal tangent frame are coordinate transformations scalars which form a covariant vector under the Lorentz group,

$$l^\alpha l^\beta \eta_{\alpha\beta} = \left(l^0\right)^2 - \left(l^1\right)^2 - \left(l^2\right)^2 - \left(l^3\right)^2.$$

Typically the time of flight is determined by measurements at points A and B. Each single point ($dl^0 = 0$) time measurement is a number $l^0$ independent of the trajectory local time scale.

Trajectory dependent infinitesimal Lorentz transformations $L_\xi(\xi)$ map the light cone subspace of the flat tangent space onto itself, producing light cone null vector deformations in accordance with eq. (3),

$$L_\xi\left(l^\alpha\right) = l^\alpha + \varepsilon \delta l^\alpha \ .$$

In order to understand the effect of the length-time scale change, the additional energy term $\varepsilon^2$ relative to an inertial system may be interpreted as a small time-relative "dressed" physical rest mass correction to the photon null mass as suggested by eq. (3). We can write a 4-velocity for an equivalent gravitational $\gamma_g$ particle on the flat tangent space. Keeping only first order terms in the limit, we obtain a curve for its velocity vector which is a quasi-null flat geodesic on the moving tangent spaces. We have an equivalent proper time $\tau$ which elapses during the $\gamma_g$ particle flight,

$$d\tau^2 \equiv -\left(\left(l^1\right)^2 + \left(l^2\right)^2 + \left(l^3\right)^2\right) + dt_c^2 \approx \varepsilon^2 dt_c^2 = \frac{\kappa dt_c^2}{2\pi} \int \frac{\sigma dV}{r} \ . \qquad (4)$$

The interval deviation, relative to a null flat geodesic, is produced by the infinitesimal curvature transformations of the moving tangent frames. This interval, invariant along the path, is a local physical time delay relative to the laboratory clock time. At time $t_c$ the $\gamma_g$ clock $\tau$ is $\varepsilon t_c$ instead of zero and clearly is a delay. The relative gravitational potential energy $\varepsilon^2$ determines this equivalent proper time delay $\delta\tau_\gamma$, which is proportional to the laboratory non inertial flight time.

## The neutrino delay.

Now we use an equivalent geometric technique based on the Ehresmann connection to obtain the delay of a hypothetical neutrino relative to the photon delay along their trajectories on the Earth geoid between the same two points. The concept of Ehresmann connections relies on replacing the infinitesimal coordinate transformations by infinitesimal group transformations. The group of the Ehresmann connection may be physically interpreted as the local tangent space curvature transformations (moving frames) or decelerations produced by the interaction energy in a non-inertial reference system.

The null geodesic in a curved space-time is the straightest null curve, and is determined by the equation for a light-like vector $l^\alpha$ whose time component $l^0$ equals $l$. The covariant derivative of this vector is determined by the gravitational Ehresmann connection [5, 6] which is a 1-form valued in the so(3,1) Lorentz group Lie algebra. The covariant derivative of the 4-velocity is the 4-acceleration. The null geodesic is

$$\frac{Dl^\alpha}{ds} = l^\mu \nabla_\mu l^\alpha = l^\mu \partial_\mu l^\alpha + l^\mu \Gamma^\alpha_{\mu\beta} l^\beta = 0. \qquad (5)$$

It is convenient to use the flying time assuming light speed $t_c$ as affine parameter instead of the zero valued interval $s$. As indicated before, the action of infinitesimal Lorentz transformations $L_\xi$ along the light trajectory is a null vector deformation. The time component $l^0$ is, up to terms of order $\varepsilon^2$,

$$l^\mu \nabla_\mu l^0 = l^\mu \partial_\mu l^0 + l^\mu \Gamma^o_{\mu\beta} l^\beta = \frac{dl}{ds} + \frac{\delta\tau}{ds} \approx \frac{dl}{dt_c} + \frac{\delta\tau}{dt_c} = 0 \ ,$$

and is related to the photon time delay $\delta\tau_\gamma$ with the same notation given in the appendix.

We recognize that the deviation of light from a flat space null geodesic, which also corresponds to the calculated deceleration or time delay, is determined by the action of the Ehresmann connection $\Gamma$ of the SO(3,1) group. We now represent this acceleration by the transformation generated by an element $a$ of the Lorentz orthogonal group Lie algebra on the null vector $l$,

$$a(\xi) \equiv l^\mu(\xi) \Gamma_\mu(\xi) \subset \text{so}(3,1) \ .$$

It is clear that $a$ generates a time delay $\delta\tau_\gamma$ which should correspond to the $\delta\tau^2$ determined by eq. (4) in terms of the constant factor $\varepsilon^2$ along the photon trajectory

$$a(l) \Rightarrow \delta\tau_\gamma = \varepsilon dt_c \ . \qquad (6)$$

When there is no gravitational field, the curve becomes a flat space-time geodesic. In these conditions the neutrino should also follow the same path of the photon. Nevertheless when there is a gravitational field we expect the neutrino spinor field $\eta$ to obey its covariant spinor equation of motion [5],

$$\overline{\sigma}^\mu \nabla_\mu \eta = \overline{\sigma}^\mu \partial_\mu \eta + \overline{\sigma}^\mu \omega_\mu \eta = 0 \ , \qquad (7)$$

where now $\omega$ is a 1-form valued in the sl(2,C) Lorentz spin group Lie algebra instead of so(3,1).

The SL(2,C) and SO(3,1) groups are 2 to 1 homomorphic. This means that the parameter space for SL(2,C) is twice the size of the parameter space for SO(3,1). A single transformation in SO(3,1) is equivalent to a pair of transformations in SL(2,C) and actually splits in two transformations: a spinor transformation and its similar conjugate transformation. Equivalently the corresponding transformation produced by the Ehresmann connection on a single spin-1/2 representation has a relative 1/2 factor with respect to the usual spin-1 vector representation transformation. We designate by $\alpha$ the sl(2,C) generator of the neutrino spinor transformation in eq. (7)

$$\alpha(\xi) \equiv \overline{\sigma}^\mu \omega_\mu(\xi) \subset \text{sl}(2,C)$$

which is isomorphic by $h$ to the $a$ so(3,1) generator of the time delay of the photon vector,

$$h(\alpha) = l^\mu \Gamma_\mu \ . \qquad (8)$$

The neutrino spinor field and its conjugate define a neutrino light-like $\lambda$ vector in space-time. The theoretical problem presented by the OPERA experiment [1] is the parallel transport of a single spinor along a physical, but geometrically unknown, space-time trajectory such as

$$\frac{D\eta}{ds} = \lambda^\mu \nabla_\mu \eta \ .$$

This spinor parallel transportation should deviate from the flat null geodesic transportation in an inertial frame. Independent of the trajectory form, the acceleration $\alpha$ generates a transformation which is similar to the one generated by $a$ and is a time dilation $\delta\tau_\nu$ by a constant factor $\varepsilon_\nu$ along the neutrino trajectory. The transformations generated by the isomorphic $a$ and $\alpha$ are related by the 2 to 1 relationship between the two Ehresmann connection group transformations. Then

$$\alpha(\eta) \Rightarrow \delta\tau_\nu = \varepsilon_\nu dt_c = \frac{\varepsilon}{2} dt_c \ . \qquad (9)$$

which shows that the deviation or delay of the neutrino from the inertial path is half the deviation or delay of the photon.

## Numerical results.

The proportionality factor $\varepsilon$ in eq, (6) which determines the photon so(3,1) transformation or delay is

$$\varepsilon^2 = \frac{\kappa}{2\pi}\int \frac{\sigma dV}{r} = \frac{4GM}{c^2 R} = \frac{4gR}{c^2} \quad ,$$

$$\varepsilon = \frac{2}{c}\sqrt{gR} = \frac{2\times\sqrt{9.8066\,m/s^2 \times 6.3781\times 10^6\,m}}{2.9979\times 10^8\,m/s} = 5.276\times 10^{-5} \quad . \quad (10)$$

The photon moving on the geodesic curve according to eqs. (1) or (5), in a non inertial laboratory system of clocks and rods, behaves as a particle which would have spent a time $\delta\tau_\gamma$ in its flight according to its own proper time. The interval $\delta\tau_\gamma$ given by eq, (6) is of the order of a small fraction of the non inertial laboratory clock flying time interval instead of exactly zero. This represents a delay in the light flying time which is unaccounted by the standard global time corrections.

The deviation or delay of the neutrino from the inertial path is half the deviation or delay of the photon,

$$\frac{d\tau_\nu}{dt_c} = \lambda^\mu \omega_\mu \eta = \frac{\varepsilon}{2} = 2.638\times 10^{-5} \quad . \quad (11)$$

The neutrino moving on its trajectory in a non inertial laboratory system of clocks and rods has a time delay $\delta\tau_\nu$ with respect to the flying time $t_c$ assuming light speed $c$. The difference of the photon and neutrino intervals, $\delta\tau_\gamma$ and $\delta\tau_\nu$, represents a gravitational delay of the photons with respect to the neutrinos.

## Applications and conclusions.

Applying the delay $\delta\tau_\gamma$ to a photon hypothetical geodesic from the CERN SPS/CNGS target to the LNGS OPERA detector, the reported experimental neutrino time anomaly [1] with respect to the photon relativistic flying time interval would be reduced by the actual photon delay to the corrected experimental value

$$\delta t' = \delta t - \delta\tau_\gamma = +2.48\times 10^{-5} t_c - 5.28\times 10^{-5} t_c = -2.8\times 10^{-5} t_c \approx \delta\tau_\nu \quad . \quad (12)$$

We find that the difference between the experimental and theoretical results, $\delta t'$ and $\delta\tau_\nu$ in eqs. (12) and (11), falls within the OPERA estimated relative errors $(\pm 0.28$ (stat.) $\pm 0.30$ (sys.)$)$ [1].

There are two UTC time-stamped measurements in the OPERA experiment, one at CERN and the other at LNGS. We apply the correction in eq. (12) to the stamped time difference because we consider that these geodesic delays are not accounted by UTC and would otherwise represent an additional uncertainty in the neutrino flying time. Corrections to the TAI clock rates have been made to account for local gravitational effects [2, 3]. It is questionable that a correction to account for this trajectory effect can be globally made within a universal time concept.

The delays are due to the equal $dl/2$ effects of length and time deformations as indicated in eqs. (1) to (3). It is possible to consider that they represent uncertainties in the definitions of a universal time or distance. The theoretical concept of time should be "the relative time" measured by an experimental observer. If the observer detects motion under an interaction, there is energy which should appear as mass [5]. The presence of this interaction energy which slows the "massless" particles is precisely what makes the system non-inertial. In general we should expect that the fastest null particle in a non-inertial system be the least interacting particle.

We may then interpret that the time anomaly of the neutrino experimental flight time by saying that the OPERA result indicates a neutrino faster than our slow hypothetical photon but that the neutrino speed is smaller than the fundamental constant $c$,

$$v_\gamma < v_\nu < c \quad . \quad (13)$$

These results also imply photon and neutrino delays of the order of 1 ns/s for every 100 m of elevation from the geoid. In addition they indicate that the neutrino trajectories are also affected by gravitational lenses but with smaller effects.

# Appendix.

The Einstein notation for the coordinates, the metric tensor and the gravitational energy tensor is [4]

$$l = ct \; ; \; x_4 = il \; ; \; ds^2 = g_{\mu\nu} dx^\mu dx^\nu \; ; \; g_{\mu\nu} = -\delta_{\mu\nu} + \gamma_{\mu\nu} \; ; \; \kappa = \frac{8\pi G}{c^2} \; ;$$

$$\gamma_{\mu\nu} = \frac{-\kappa}{4\pi} \int \frac{T^*_{\mu\nu}}{r} dV \; ; \; \gamma_{11} = \gamma_{22} = \gamma_{33} = \frac{-\kappa}{4\pi} \int \frac{\sigma dV}{r} \; ; \; \gamma_{44} = \frac{+\kappa}{4\pi} \int \frac{\sigma dV}{r}$$

Modern differential geometry avoids the use of general coordinates by projecting the coordinate 1-forms onto Cartan's orthonormal moving frames [7] obtaining a set of coordinate scalars which form a covariant vector under the Lorentz group.

$$\theta^\alpha = \theta^\alpha_\mu dx^\mu \; ,$$
$$g_{\mu\nu} = \eta_{\alpha\beta} \theta^\alpha_\mu \theta^\beta_\nu \; .$$

The SO(3,1) group transformation which leaves invariant the metric $\eta$ on the flat tangent space is a Lorentz transformation. It is not a coordinate transformation but determines an Ehresmann connection which is a sequence of group actions along a trajectory.